# Robust Sequential Steady-State Analysis of Cascading Outages

Amritanshu Pandey *[1,2], Aayushya Agarwal *[1], Marko Jereminov[1], Martin R. Wagner[1], David M. Bromberg[2], Larry Pileggi[1]

[1]Dept. of Electrical and Computer Engineering
Carnegie Mellon University
Pittsburgh, PA, USA

[2]Pearl Street Technologies
Pittsburgh, PA, USA

*Abstract-* Simulating potential cascading failures can be useful for avoiding or mitigating such events. Currently, existing steady-state analysis tools are ill-suited for simulating cascading outages as they do not model frequency dependencies, they require good initial conditions to converge, and they are unable to distinguish between a "collapsed grid state" from a "hard-to-solve test case." In this paper, we extend a circuit-theoretic approach for simulating the steady-state of a power grid to incorporate frequency deviations and implicit models for under-frequency and under-voltage load shedding. Using these models, we introduce a framework capable of robustly solving cascading outages of large-scale systems that can also locate infeasible regions. We demonstrate the efficacy of our approach by simulating entire cascading outages on an 8k+ nodes sample testcase.

## I. INTRODUCTION

Modern society depends on the secure and reliable operation of the electric grid. Cascading outages represent a class of events that can significantly impact the electric grid and create wide-spread socio-economic damages. The North American Electric Reliability Corporation (NERC) defines cascading outages "as the uncontrolled loss of any system facilities or load, whether because of thermal overload, voltage collapse, or loss of synchronism, except those occurring as a result of fault isolation" [1].

NERC has released several guidelines to mitigate the likelihood of cascading outages occurring in the grid. Amongst those, NERC standards TPL-001-4 [2] and CIP-014-2 [3] require evaluation of the impact of extreme contingencies that may cause cascading outages. On the operations side, Emergency Operations EOP-003-1 [4] requires that: "After taking all other remedial steps, a Transmission Operator or Balancing Authority operating with insufficient generation or transmission capacity shall shed customer load rather than risk an uncontrolled failure of components or cascading outages of the Interconnection" by implementing Special Protection Systems (SPS) and other routines to automatically shed load under adverse events.

Therefore, to properly analyze cascading outages, a simulation framework must (i) solve extreme contingency cases from initial conditions that are far from the solution; (ii) identify and locate collapsed (infeasible) grid locations; and (iii) include frequency state into its framework to model the impact of generator droop control and automatic protection schemes such as frequency dependent load shedding. A framework capable of satisfying these requirements will allow planning engineers to distinguish a cascaded grid scenario from a divergent scenario. Furthermore, such a framework would be able to converge infeasible test cases (i.e., cases operating beyond the tip of the nose curve) and allow planning engineers to locate weak sections of the grid. Additionally, it is also important for the framework to robustly simulate any remedial actions in order to accurately analyze the grid during a cascade outage. These include Under-Frequency Load Shedding (UFLS) and Under Voltage Load Shedding (UVLS) schemes.

Existing frameworks for simulating cascading outages have tried to incorporate these features in both sequential power flow analysis [5], [6] as well as transient analysis [7]-[8]. In general, transient analysis is slow and is therefore only performed for critical contingencies in the system. A sequential steady-state power flow analysis offers runtime advantages to study a broad range of outages. However, existing steady-state tools in the industry and academia do not satisfy the previously stated requirements of incorporating frequency information and/or solving infeasible cases. This is highlighted in a recent report [9] by the Task Force on Understanding, Prediction, Mitigation, and Restoration of Cascading Failures that stated that "the tools for directly assessing and mitigating large cascading failures are not yet well developed."

Overall, there has been significant research on addressing key elements required to develop a robust tool for cascading analysis. [10]-[11] have improved the robustness of convergence of complex and large "hard-to-solve" cases by incorporating limiting and homotopy methods. Modeling the frequency state is also broached by many existing works [11]-[13]; however, these approaches use outer loops to resolve discontinuous models which can cause simulation convergence issues. Furthermore, detection and localization of infeasible grid states is also ongoing work [14]-[15]. Continuation power flow [16] was previously proposed to solve infeasible test cases (operating beyond the tip of the nose curve), but requires solving the base case first, which itself is hard to achieve for complex, large-scale test cases when the initial condition is far from the solution. Other optimization-based methodologies [15] have also attempted to solve infeasible test cases, but they generally suffer from a lack of convergence robustness and have only been tested on small, well-conditioned networks.

The current-voltage state variables-based power flow formulation in [11] proposes to model the transmission and distribution grid networks as equivalent circuits. This approach has been demonstrated to enable robust convergence of complex transmission or distribution networks via use of circuit simulation methods [11]. Recently, we have expanded this simulation framework to further solve infeasible test networks and locate weak system areas that would correspond to a collapsed grid state [14]. In this paper, we expand the circuit-based formulation to model frequency deviations while implicitly capturing UFLS and UVLS without the need for outer loops. This enables us to develop a framework to accurately simulate cascading outages while localizing and identifying collapsed sections of the grid.

## II. Background

### A. Governor Power Flow

Previous works [11]-[13] have explored the use of a frequency variable in power flow formulations, including optimization-based methods [5]. These frameworks generally adjust the real power of the generator based on the change of the grid's frequency $\Delta f$:

$$P_G = P_G^{SET} - \frac{P_R}{R}\Delta f \qquad (1)$$

where, $P_G$ is the frequency-adjusted real power output of the generator, $P_G^{SET}$ is the set generator real power, and $P_R$ and $R$ define the primary frequency response of the generator based on droop control and inertia. However, these methods typically account for real power limits of the generator in the outer loop of the solver, thereby causing significant convergence issues due to piecewise discontinuous modeling [17], where:

$$P_G = \begin{cases} P_G^{MAX}, & \text{if } P_G > P_G^{MAX} \\ P_G^{MIN}, & \text{if } P_G < P_G^{MIN} \\ P_G^{SET} - \frac{P_R}{R}\Delta f, & \text{otherwise} \end{cases} \qquad (2)$$

In [10] and [17], we have shown that convergence issues due to piecewise discontinuous modeling of limits can be avoided by using implicit models for system control including real power control of a generator based on frequency deviations. It was shown that a generator's change in real power due to the primary response can be captured by formulating the change in real power as a function of frequency including real power limits [18]. This continuous model implicitly captures the generator real power limits without the need for outer iteration loops.

### B. Current-Voltage Approach to Power Flow Analysis

Current-voltage (I-V) based formulations have been explored in the past for performing power flow and three-phase power flow analyses [11], [19]. Amongst these approaches, a recently introduced equivalent circuit approach [11] maps the different network models of the grid (e.g. PV, PQ etc.) into their respective equivalent circuits and further aggregates them together to create the whole network model to solve for the node voltages and branch currents. This approach has also been previously extended with circuit simulation methods to preserve the physical behavior of the grid elements by avoiding solutions that include low-voltages [11] and generators operating in an unstable region [17]. Robust convergence to meaningful solutions from initial conditions that are far from the solution is essential for simulating cascading outages, and we demonstrate the efficacy of an extension of this circuit-theoretic simulation framework for such simulation problems in this paper.

### C. Optimization Using Equivalent Circuit Approach

The equivalent circuit approach described above can be extended to formulate constrained optimization problems via the use of adjoint theory [14], [20]. In this approach, in addition to the power flow circuits, adjoint circuits are added to the framework [14] to represent the necessary first-order optimality conditions of the optimization problem. The solution of this net aggregated circuit (power flow circuits and adjoint circuits) corresponds to an optimal solution of the optimization problem and is obtained by using the circuit simulation techniques described in [11], [21].

For simulation of cascading outages, it is essential to differentiate "hard-to-converge" grid-state from "infeasible" grid-state. Therefore, the optimization problem in [14] that was previously proposed to detect infeasibility in regular power flow test cases by introducing an *infeasibility current source* $I_F$ at each bus, is extended to include frequency control and special protection schemes (UVLS and UFLS) to simulate cascading outages, as shown in later sections of this paper.

## III. Implicit Model for Load Shedding

In order to simulate the effect of a contingency during a cascade, it is important to consider automatic schemes designed to protect the stability of the grid. Under Frequency Load Shedding (UFLS) and Under Voltage Load Shedding (UVLS) are a set of control mechanisms designed to restore and maintain frequency and voltage stability respectively, by disconnecting loads to match the supply of generators [22]. Both these load shedding schemes can shed load discretely or continuously (as described in [23]) by disconnecting parts of loads based on the frequency of the grid or the voltage at the bus.

In both discrete and continuous load shedding mechanisms, load shedding only occurs once the frequency of the grid or the voltage at the bus is below a certain threshold. To study the steady-state effects of UFLS and UVLS, we incorporate this behavior into power flow and extend the PQ load model as shown in (3)-(4). This facilitates both discrete and continuous load shedding.

$$P_L = P_L^{set}(1 - \alpha^{UFLS})(1 - \alpha^{UVLS}) \qquad (3)$$
$$Q_L = Q_L^{set}(1 - \alpha^{UFLS})(1 - \alpha^{UVLS}) \qquad (4)$$

The new variable $\alpha = \{\alpha^{UFLS}, \alpha^{UVLS}\} \in [0,1]$ describes load shedding due to UFLS or UVLS at the steady-state grid response. When $\alpha = 0$, the load experiences no load shedding ($P_L = P_L^{set}$ and $Q_L^i = Q_L^{i\,set}$), whereas, $\alpha = 1$ represents full load shedding ($P_L^i = P_L^{i\,set}$ and $Q_L^i = Q_L^{i\,set}$).

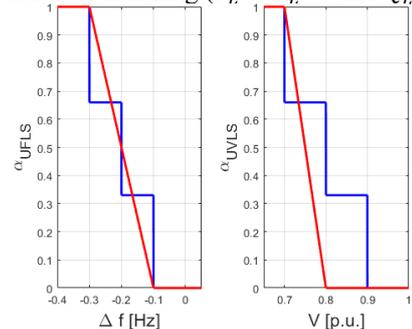

Figure 1: Discrete (blue) and Continuous (red) UVLS and UFLS.

In order to model discrete load shedding during power flow, a PQ load at a distribution station is segmented into separate PQ loads, each corresponding to a part of the load being shed at a particular threshold frequency or voltage. Each segmented load then has a corresponding $\alpha$. For a continuous load shedding scheme, the entire PQ load can be modelled as a whole with a single $\alpha$.

For each load, $\alpha$ is controlled by the frequency of the grid or the voltage at the bus depending on the scheme, described in the following sections.

### D. Implicitly modeling UFLS

Under the UFLS scheme, $\alpha^{UFLS} \in [0,1]$ is controlled by the frequency of the grid. Ideally, in a discrete load shedding scheme:

$$\alpha^{UFLS} = \begin{cases} 0, & \Delta f > f_{SET} \\ 1, & \Delta f \leq f_{SET} \end{cases} \quad (5)$$

However, (5) is a discontinuous function and causes numerical issues during Newton-Raphson (NR). In order to model the discontinuous behavior implicitly without any outer loop, we approximate the behavior of UFLS in (5) by a continuous differentiable function (6) and is depicted in Fig. 2.

$$\alpha^{UFLS} = \begin{cases} 0, & \text{Region 1} \\ a_{\min}\Delta f^2 + b_{min}\Delta f + c_{min}, & \text{Region 2} \\ -\beta(\Delta f - f_{SET}), & \text{Region 3} \\ a_{max}\Delta f^2 + b_{max}\Delta f + c_{max}, & \text{Region 4} \\ 1, & \text{Region 5} \end{cases} \quad (6)$$

(6) is a smooth function with a continuous first derivative that inherently considers the threshold frequency ($f_{SET}$) as well as bounds $\alpha^{UFLS} \leq 1$. Region 1 in (6), also shown in Fig. 2, represents no load shedding when the frequency of the grid is above $f_{SET}$. In order to model discrete load shedding step at $f_{SET}$, the model moves from Region 1 to Region 5 through continuous sections. It is apparent from Fig. 2 that in order to tightly match the discrete step, $\beta$ must be large (> 100).

A model for continuous load shedding is described by:

$$\alpha^{UFLS} = \begin{cases} 0, & \Delta f > f_{SET} \\ -K(\Delta f - f_{SET}), & \Delta f \leq f_{SET} \end{cases} \quad (7)$$

where, $K$ is a predefined factor that controls the amount of load shedding at each frequency deviation. However, (7) is also a discontinuous function that can cause numerical stability issues. As a result, (6) can also be used to model the continuous load shedding by equating the factor $\beta$ to $K$.

Regions 2 and 4 are vital quadratic patching functions for (6) that provide first derivative continuity, thus improving simulation convergence properties. Before the simulation begins, we solve for the quadratic parameters by matching their values and the first derivative of the quadratic function to the adjacent functions at the intersection points. This forces the function to be continuous and smooth for all frequencies, thereby improving robustness during simulation as it does not have the deficiencies of discontinuous piecewise models. In using the optimization framework described in Section II.C, it is also necessary to calculate the search direction using the second derivative of the implicit models. The model given in (6) is first-order continuous and its second order derivative is estimated as done in [17] to provide a reasonable approximation that allows the model to converge toward a feasible solution.

While modelling a discontinuous load shedding scheme using the implicit UFLS model, the framework may converge to a solution with $0 < \alpha^{UFLS} < 1$, i.e. within a region that is not fully connected nor fully disconnected. In this case, we are able to apply an outer loop to snap the function of $\alpha^{UFLS}$ to the closer bound (0 or 1). In practice, this outer loop does not result in convergence issues because the previous solution is close enough to the true solution of the grid to allow fast convergence using N-R.

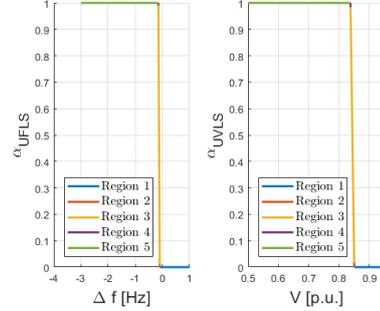

Figure 2: Continuous implicit models for UVLS and UFLS.

### E. Implicit Model of UVLS

UVLS behaves similarly to the UFLS scheme with the exception that $\alpha^{UVLS}$ is controlled by the bus voltage $V$. UVLS schemes typically disconnect parts of a load if the bus voltage is below a certain minimum threshold, $V_{SET}$. The steady-state of UVLS can be described as:

$$\alpha^{UVLS} = \begin{cases} 0, & V > V_{SET} \\ 1, & V \leq V_{SET} \end{cases}$$

Similar to the ideal UFLS model, however, equation (8) is a discontinuous function that cannot guarantee convergence using numerical methods such as NR. To remediate this, we apply the same model as UFLS, but with the controlling variable as voltage (V), as shown in Fig. 2.

$$\alpha^{UVLS} = \begin{cases} 0, & \text{Region 1} \\ a_{\min}V^2 + b_{min}V + c_{min}, & \text{Region 2} \\ -\beta(V - V_{SET}), & \text{Region 3} \\ a_{max}V^2 + b_{max}V + c_{max}, & \text{Region 4} \\ 1, & \text{Region 5} \end{cases} \quad (8)$$

Like the implicit model for the discrete UFLS, we use an outer loop to handle simulations where the solution converges to $0 < \alpha^{UVLS} < 1$. In practice, this outer loop also does not result in any convergence issues.

## IV. PROBLEM FORMULATION

In this section, we develop a framework for simulating cascading outages based on a robust current-voltage formulation [14], [10]. This algorithm, depicted in Fig. 3, extends the power flow formulation by incorporating under frequency and voltage load-shedding and is able to identify if the network has partially or totally collapsed.

### A. Cascade Analysis Stage 1 Module

The first stage of the framework solves for the steady state of the grid under a contingency using power flow including UFLS and UVLS schemes. Mathematically, stage I can be described as an optimization problem:

$$\min_X \quad \|I_F^R, I_F^I\|_2^2 \quad (9a)$$
$$\text{s.t.} \quad YV - I_F = 0 \quad (9b)$$
$$P_S^i + \Delta P_S^i = V_R^S I_R^S + V_I^S I_I^S, i \in \text{slack} \quad (9c)$$
$$\Delta P_S^i - \frac{P_R^i}{R_i}\Delta f = 0, i \in \text{slack} \quad (9d)$$

$$P_G^i = P_G^{i,SET} + \Delta P_G^i, i \in PV, \quad (9e)$$

$$\Delta P_G^i - \langle \frac{P_R^i}{R_i} \Delta f \rangle = 0, i \in PV \quad (9f)$$

$$P_L^i = P_L^{i,set}(1 - \alpha^{UFLS})(1 - \alpha^{UVLS}), i \in PQ \quad (9g)$$

$$P_L^i = Q_L^{i,set}(1 - \alpha^{UFLS})(1 - \alpha^{UVLS}), i \in PQ \quad (9h)$$

where $\langle x \rangle$ operator bounds $x$ in the predefined range given by $x_{min}$, and $x_{max}$ and is modeled by the implicit continuous model for saturating real power given in [17]. (9a) is the objective function that minimizes the real and reactive feasibility currents $I_F$ that are added to each node ($v \in \mathcal{V}$) in the system. For a feasible system state, these currents are minimized to zero, whereas under an infeasible system state, the currents have a non-zero magnitude at locations that lack either real or reactive power, corresponding to a collapsed system. (9b) represents the non-linear network Kirchhoff Current Law constraints for each node $v$ in the system. (9c) and (9d) correspond to the output of the slack generator ($P_S^i$) in the set of slack buses (slack). Similarly, (9e) and (9f) represent the implicit modeling of AGC/droop control as a function of frequency for all generators in the set of $PV$ buses, and (9g) and (9h) represent the implicit modeling of UVLS and UFLS for all loads in the set of $PQ$ loads.

The problem described above is a non-convex quadratic programming problem and is solved in our approach using Newton's method [14] to obtain a local optimum solution. We also make use of circuit-based limiting and homotopy methods in [11] ensure a coherent optimal solution while also pushing the solution away from low-voltage solutions.

### B. Incorporating protective relaying data for Stage II

After finding an optimal steady-state solution in Stage I, we must report any device in the grid that has hit its limit and deactivate it prior to simulating the next cascading outage stage. In this framework, we incorporate protective relaying limits for line and transformer overloads as well as generator voltage limits. The remedial action schemes, such as UVLS and UFLS, were previously incorporated in Stage I.

### V. SIMULATION ALGORITHM

The overall algorithm for simulating cascading outages in circuit-theoretic approach, with implicit modeling of UFLS and UVLS schemes and frequency state, is shown in Fig. 3. The algorithm begins by taking a network, with an initiating event, that will be simulated to study the cascading effect. We also create a set X, initialized as empty, to represent the lines, transformers and generators that have tripped due to over-loading and exceeding over-voltage limits. The Stage I of the problem as formulated in Section IV.A is solved with elements in set X de-activated and simulates the network case while modelling the frequency state and special protection systems (UFLS and UVLS). If, after running Stage I of the solver, the network case (or a subset of islands) is found infeasible (operating past the tip of the nose curve), then the localized areas responsible for collapsing the grid is reported to the user. If the output of the Stage I solver is found to be feasible for any island (i.e. converges to a solution with $I_F = 0$), the solver adds the lines, transformers, and generators that have violated their limits in the feasible islands to the set X. The devices in set X are then tripped and the algorithm is re-run until no device is found violating its limit (set X is empty), or the case (all islands) is found to be infeasible (i.e. converges to a solution with non-zero $I_F$), representing a system wide blackout. The output for the simulation then reports the state of each island in the network case as either secure or collapsed.

The algorithm described for simulation of a cascading outage is fully parallelizable for each contingency without any co-dependencies. Therefore, in order to analyze a set of high-risk contingencies, they can be solved in parallel with ease.

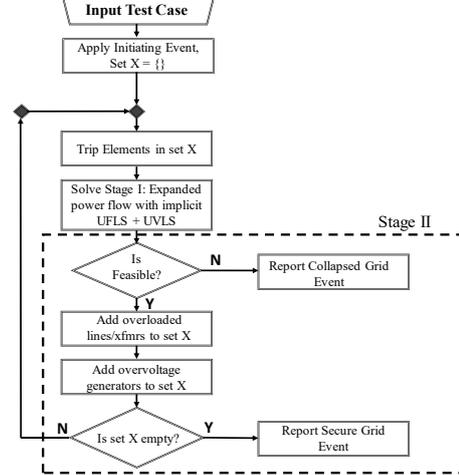

*Figure 3: Flowchart for implementation circuit theoretic cascading analysis.*

### VI. RESULTS

Here we demonstrate the advantages of our approach for simulating cascading outages on a simple 14-bus system and a larger 8k+ node system. For all test cases, we incorporate frequency information as well as implicit models for UFLS and UVLS. Power flow is analyzed by our prototype simulation tool, SUGAR[10].

### A. Advantages of Implicit UFLS and UVLS models

Using the standard IEEE14 bus test case with loading factor of 4.2 (LF) (load parameters P and Q and generator P scaled), we completed simulation runs with an (i) implicit UFLS model as described in Section III and (ii) the UFLS model in the outer loop as described in [5]. The UFLS model is described by following parameters:

| Frequency Threshold ($f_{MIN}$) | % load shedding |
|---|---|
| -0.3 | 17% |

TABLE 1: COMPARISON OF RESULTS BETWEEN IMPLICIT UFLS MODEL AND UFLS MODEL IN OUTER LOOP.

| | Case Name | Loading Factor | Solution Achieved? | Δf (hz) |
|---|---|---|---|---|
| Implicit UFLS | IEEE-14 | 4.2 | Converged | -0.476 |
| UFLS in outer loop | IEEE-14 | 4.2 | Did not converge* | NA |

*Inner loop failed to converge in 100 iterations

The results shown in Table 1 demonstrate that convergence of the UFLS model based on an outer loop is highly dependent on the inner loop convergence. However, if the inner loop of the network is infeasible (as is the case with 4.2 LF without load shedding), it prevents system convergence. With an implicit UFLS model, however, the UFLS control is solved concurrently in the inner loop, and convergence is easily achieved.

### B. Steady-State Simulation of Cascading Outage

The framework shown in Fig. 3 allows us to sequentially simulate the cascading outage on a modified 8k+ node system [24] while considering the steady state effects of UFLS, UVLS and generator frequency control mechanisms. The parameters are shown in Table 2.

TABLE 2: FREQUENCY RESPONSE PARAMETERS.

| UFLS ($\Delta f_{SET}$) | UVLS ($V_{SET}$) | Primary Control ($\frac{P_R}{R}$) |
|---|---|---|
| -0.3 Hz | 0.95 p.u. | 0.1 MW/Hz |

The simulation results in Fig. 4 identify the devices that were tripped in each cascading stage, as well as show the islands created and their respective frequencies. The simulation completed by creating 3 islands, all of which collapsed, creating a system wide blackout, with a steady state frequency deviation of the infeasible main island as -0.016Hz. By using implicit models and circuit heuristics, the framework robustness was demonstrated for this case. The infeasibility is indicated by the non-zero infeasibility currents in cascading stages 2 and 3. The final stage results in a system wide black out due to a collapse of all islands.

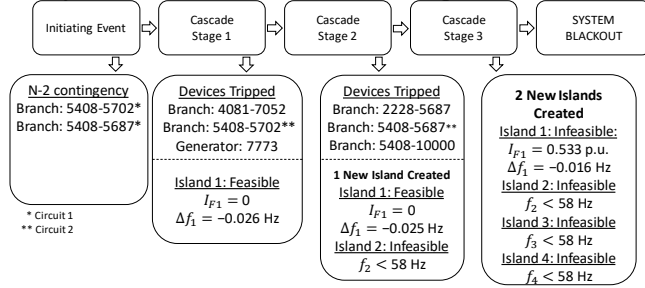

*Figure 4: Results of cascading simulation on 8k+ node system.*

Fig. 5 demonstrates the infeasibility currents in different areas of the system during the final stage. It is visible from the bar plot that areas around 48 through 55 are the weakest regions with the largest infeasibilities and one of the primary causes of the cascading outage.

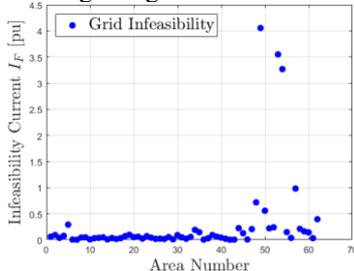

*Figure 5: Infeasibility currents during system blackout stage.*

## VII. CONCLUSIONS

In this paper, we developed a robust framework to simulate the steady-state of sequential cascading outages. To achieve this, we extended the current-voltage based power flow formulation to model frequency and developed implicit models for under frequency load shedding and under voltage load shedding schemes. These models were shown to have better convergence characteristics than discontinuous outer-loop models of UFLS and UVLS. Importantly, the optimization used to solve for the steady-state with minimization of "feasibility" currents enables simulation of large cascading outages, such as the 8k+ bus network, while localizing and identifying any collapsed sections of the grid.

ACKNOWLEDGMENT

This work was supported in part by the Defense Advanced Research Projects Agency (DARPA) under award no. FA8750-17-1-0059 for RADICS program, and the National Science Foundation (NSF) under contract no. ECCS-1800812.